\documentclass[pra,aps,groupaddress,showpacs,prd,preprint]{revtex4-1}
\usepackage{epsfig,amsmath,graphicx,amssymb}
\def\be{\begin{equation}}
\def\ee{\end{equation}}
\def\bea{\begin{eqnarray}}
\def\eea{\end{eqnarray}}
\def\bes{\begin{subequations}}
\def\ees{\end{subequations}}

\newcommand{\fr}{ \frac}

\begin{document}
%%%%%%%%%%%%%%%%%%%%%%%%%%%%%%%%%%%%%%%%%%%%%%%%%%%%%%%%%%%%%%%%%%%%%%%%%%%%%%%%%
\title{Dark solitons dynamics and snake instability in superfluid Fermi gases trapped by an anisotropic harmonic potential}
\author{Wen Wen and Changqing Zhao}
\affiliation{Department of Mathematics and Physics, Hohai
University, Changzhou Campus, Changzhou 213022, China}
\author{Xiaodong Ma}
\affiliation{College of Physics and Electronic Engineering, Xinjiang Normal University,
Urumchi 830054, China}
%
%%%%%%%%%%%%%%%%%%%%%%%%%%%%%%%%%%%
\date{\today}

%%%%%%%%%%%%%%%%%%%%%%%%%%%%%%%%%%%

\begin{abstract}
We present an investigation of generation, dynamics and stability
of dark solitons in anisotropic Fermi gases for a range of particle
numbers and trap aspect ratios within the framework of the order-parameter equation.
We calculate the periods of dark solitons oscillating in a trap, and
find a good agreement with the results based on the Bogoliubov-de Gennes
equations. By studying the stability of initially off-center dark solitons under various tight
transverse confinements in the unitarity limit, we not only give the criterion of dynamical stability,
but also find that the soliton and a hybrid of solitons and vortex rings can be characterized
by different oscillation period. The stability criterion is not fulfilled by the parameters of the recent
experiment [Nature {\bf 499}, 426 (2013)]. Therefore, instead of a very slow oscillation as observed
experimentally, we find that the created dark soliton undergoes a transverse snake
instability with collapsing into vortex rings, which propagate in soliton-like manner
with a nearly two times larger period.
\end{abstract}

\pacs{03.75.Ss, 03.75.Lm}

\maketitle

%%%%%%%%%%%%%%%%%%%%%%%%%%%%%%%%%%%%%%%%%%%%%%%%%%%%%%%%%%%%%%%%%

\section{Introduction}
Dark solitons, namely localized density dips with a phase-jump
across their density minimum, are the most fundamental nonlinear
excitations in nonlinear dispersive media. They appear in many areas
of science, such as water waves, nonlinear optics, biophysics and
plasma and particle physics\cite{dau}, and more recently in
Bose-Einstein condensate\cite{pet,kev,fra}. Since the first
observation of the crossover from a Bardeen-Cooper-Schrieffer(BCS)
superfluid to a Bose-Einstein condensation(BEC) in ultracold
fermionic atomic gases\cite{hara,reg,abr,zwi,gio}, understanding
formation, dynamics and stability of solitons in a strongly
interacting fermionic system has been attracted great
attention\cite{yef,ada}, and explored
theoretically\cite{ant,spu,sco,ren,sco1,bul,wen2,khan}.

Based on the BCS mean-field theory\cite{pdg}, the existence and
properties of black solitons in the BCS-BEC crossover were
demonstrated\cite{ant} by the real solutions of the Bogoliubov-de
Gennes(BdG) equations. The more general case of the complex
solutions corresponding to grey solitons was also
considered\cite{spu,sco1}. Furthermore, the periodic dynamics of
dark solitons in a harmonic trap\cite{sco,ren} and two solitons
collision\cite{sco1} were studied by numerically solving the
time-dependent BdG equations. It was predicted that the oscillation
period of a soliton in a harmonic trap increases as one moves from
the BEC to BCS regimes\cite{sco,ren}. From the computational side,
calculating solutions to both the time-independent and
time-dependent BdG equations is numerically intensive, since they
require a self-consistent calculations of single-particle states
whose number increases linearly with the number of particles. For
this reason, these investigations have essentially been restricted
to Fermi gases confined in a box\cite{ant,spu} or one dimensional(1D)
trapping potential\cite{sco1,ren} and for a small number of atoms,
that is essentially quasi-1D. However, it is not particularly
relevant to current experimental setting.

In the very recent MIT experiment performed by Yefsah {\it et al.} with
a fermionic superfluid of $^{6}$Li near a Feshbach resonance the
long-lived solitons were observed\cite{yef}. These authors created
dark solitons by phase imprinting in the cigar-shaped superfluid.
Instead of {\it in situ} imaging solitons at the Feshbach resonance,
the visualization of solitons has been relied on the method of
time-of-flight. It was obtained after releasing the superfluid cloud
from its trap and letting it expand with the rapid ramp to the
weakly interacting BEC regime. The oscillation period of dark
solitons in the unitary regime was measured, which is ten times
larger than one predicted by the BCS mean-field theory\cite{sco,ren}.
Note that Yefsah {\it et al.} essentially have prepared the
superfluid Fermi gases containing about $2\times10^5$ atom pairs,
and confined in an external anisotropic harmonic trap, which is not
satisfied by the quasi-1D condition. Hence it is of great interest in
generation, dynamics and stability of dark solitons in a genuinely
three-dimensional(3D) superfluid Fermi gas.

Different from previous investigations by the extend BdG
equations\cite{ant,spu,sco,ren,sco1}, our theoretical investigations
are based on the time dependent order-parameter
equation\cite{sal,wen3}. The order-parameter equation only can
describe superfluidity features macroscopically, but its
mathematical framework is simple that involves a single function of the
coordinate, i.e. the superfluid density. Thus we encounter no
limitation in the number of particles and external potentials, and
analyze easily and clearly. Our calculation is carried out for a
wide range of the number of atoms and trap aspect ratios.
We not only present the oscillation periods of dark soliton in the
trapped Fermi gases containing small number of atoms, but also find that
as the number of atoms is increased by two orders, the period of stable solitons
increases 8\%. By examining the effects of transverse confinements on the stability of
initially off-center solitons through their phase profiles, we give the criterion
of dynamical stability. Finally we find that the dark soliton created in the MIT
experiment is subject to snake instability, splitting into two vortex rings, and eventually reduces to one vortex ring, which performs a soliton-like oscillatory motion.

This paper is organized as follows: after a brief
description of the order-parameter equation and numerical method in Sec.II,
the dynamics of dark solitons in the quasi-1D regime is studied in Sec.III,
and the soliton periods along the BCS-BEC crossover are compared with ones obtained
from the BdG equations. In Sec.IV, the snake instability of dark soliton in the unitary
limit is studied, and the dynamic stability criterion is given. The results of our calculation
on the dark soliton dynamics in the parameters of the MIT experiment are presented in
Sec.V. Finally a conclusion is given in Sec.VI.

\section{Model and method}
We consider an ultracold Fermi gas at zero temperature, in which
fermionic atoms have two spin states with equal number. In a ground
state all atoms are paired and in the superfluid state, which can be
described by the following time dependent order-parameter (or
macroscopic wavefunction)
equation\cite{kim,zub,sal,wen3,adh,adh1,anc,rup,xio} \be\label{gg}
i\hbar\fr{\partial \Psi_s}{\partial
t}=\left[-\fr{\hbar^2\nabla^2}{2M}+V_s({\bf
r})+\mu_s(n_s)\right]\Psi_s,
\ee
where $\Psi_s$ is the order parameter of fermionic atomic pairs in
the superfluid state, with the superfluid density $n_s=|\Psi_s|^2$
and the normalized condition $\int d{\bf r} |\Psi_s|^2=N$ ($N$ is
the total atomic pair number of the superfluid Fermi gas, also equal
to the number in each spin state). $M$ is the mass of atom pair
(i.e. $M=2m$ with $m$ being atomic mass), and $V_s({\bf r})$ is the
external potential. We write the order parameter
$\Psi_s=\sqrt{n_s}e^{i\Phi_s}$ in terms of its amplitude
$\sqrt{n_s}$ and the phase $\Phi_s$\cite{pet}, which can be
understood by the superfluid velocity ${\bf v}_s({\bf r}, t)=\hbar\nabla\Phi_s({\bf r},
t)/M$\cite{pet,wen3}.

Defining a dimensionless interaction parameter
$\eta\equiv1/(k_Fa_s)$, where $k_F$ is the Fermi wavenumber and $a_s$
is $s$-wave scattering length, one can distinguish several different
superfluid regimes: BCS regime($\eta\leq-1$), BEC
regime($\eta\geq1$), and BCS-BEC crossover regime($-1<\eta<1$). A
special case $\eta=0$ is called unitarity limit where the scattering
length is infinity.  The BCS-BEC crossover regime is a strongly
interacting regime, and BCS($\eta\ll-1$) and BEC($\eta\gg 1$) limits
are actually weakly interacting. In general, the expression for the
equation of state $\mu_s(n_s)=2\mu(n)$ with $n=2n_s$ being atomic
density is very complicated,  but it can be fitted by the analytical
formula based on the Monte Carlo data\cite{geas}, and approximated by the
polytropic approximation\cite{dia,mann,wen1}
\begin{subequations}
\label{poly}
\bea
& &  \label{poly1} \mu_s(n_s)=2\mu^0 \left(n_s/n^0\right)^{\gamma},\\
& &  \label{poly2} \gamma=\gamma (\eta)=\fr{n}{\mu}\,\frac{\partial
\mu}{\partial n}= \fr{\fr{2}{3}\sigma
(\eta)-\fr{2\eta}{5}\sigma^{\prime }(\eta)+
\fr{(\eta)^{2}}{15}\sigma^{\prime \prime}(\eta)}{\sigma (\eta)-
\frac{\eta}{5}%
\sigma^{\prime }(\eta)},
\eea
\end{subequations}
where $\mu^0$ and $n^0$ are respectively reference chemical
potential and particle number density\cite{wen1}.  Usually we take
reference particle number density
$n^0=(2mE_F)^{3/2}/(6\pi^2\hbar^3)$ to be the per spin density of
the non-interacting Fermi gas at the trapping center,
and reference chemical is thus $\mu^0=E_F
(\sigma(\eta)-\eta\sigma^{\prime}(\eta)/5)$ proportional to the
Fermi energy $E_F=(\hbar k_F)^2/(2m)$. The order-parameter equation
incorporated with the equation of state allows one to investigate
the smooth crossover from the BEC limit to BCS regime\cite{noz}
in a unified way.

It is noticed that in the unitarity limit ($\eta=0,\gamma=2/3$), the
order-parameter equation is exactly equivalent to one derived by
Salasnich {\it et al.} from an extended Thomas-Fermi density
functional theory\cite{sal}. In the BEC limit ($\eta\gg 1,
\gamma=1$), the order-parameter equation coincides exactly with one
derived by Pieri and Strinati based on BdG equations\cite{pie}. It
has been demonstrated that the order-parameter equation is very
reliable to capture ground-states properties\cite{dia,adh} and
low-energy collective dynamics\cite{adh,wen1} in the BCS-BEC
crossover. Furthermore, the results given by the order-parameter
equation in the BEC side of the crossover are found to be in
good agreement with the zero-temperature BdG
equations\cite{lsa,anci}. However, the order-parameter equation can
not completely capture dynamical properties in the BCS
regime($\eta<0$) for the reason that dynamical behaviors can easily
result in pair breaking due to very small gap energy\cite{zai}, while the
order-parameter equation ignores single atom excitation.

We consider a cylindrically symmetric harmonic trap
\be\label{exp}
V_s({\bf r})=\fr{1}{2}M\omega^2_z(\lambda^2r^2+z^2),
\ee
where $(r,z)$ are cylindrical coordinates with $r=\sqrt{x^2+y^2}$.
The aspect ratio (anisotropy) of the trap is defined by $\lambda=\omega_{\perp}/\omega_z$,
with the trapping frequencies $\omega_{\perp}$ and
$\omega_z$. So the Fermi energy for the
fermions trapped by a 3D harmonic potential is given by
$E_F=\hbar(6N\omega^2_{\perp}\omega_z)^{1/3}$. The
axial trapping frequency is $\omega_z=2\pi\times 10.66\text{Hz}$ in the MIT experiment\cite{yef}, and
the transversal frequency $\omega_{\perp}=\lambda\omega_z$ is determined by the fixing
axial frequency and different aspect ratios. We choose total
number of atom pairs in a wide range $N=2\times 10^5\sim 2\times
10^2$. In the following, the length is in units of
$a_z\equiv\sqrt{\hbar/(M\omega_z)}=8.89\rm{\mu m}$ and time is in
units of axial trapping period $T_z=93.76\text{ms}$\cite{yef}. Density profiles are
presented by normalized cross-sectional densities at the $y=0$
plane, i.e. $n_s(x,z)=\fr{a^3_z}{2\sqrt{2}N}|\Psi_s(x,y=0,z)|^2$.

Recently we have presented the dark soliton solutions of quasi-1D
order-parameter equation by the multiple
scale method in the small-amplitude limit\cite{wen2}. Later bell
solitons along the BCS-BEC crossover as exact soliton solutions of
the order-parameter equation in arbitrary amplitudes were found
analytically\cite{khan}. Dark solitons have been
created in Fermi gases by using phase imprinting technique\cite{yef},
which originally has been proposed to generate vortices and solitons
in weakly interacting atomic BECs\cite{dob}, and experimentally
implemented\cite{bur,den}. The main ideal of this technique is
described as shining an off-resonance laser on a condensate in order
to create phase steps between its different parts. Instead of
exposing the analytic 1D soliton solution to 3D, we
simulate the phase imprinting method to generate initial solitons in
anisotropic Fermi gases along the BCS-BEC crossover.

We solve the order-parameter equation by discretizing with the
split-step Crank-Nicolson algorithm\cite{pmu}. The initial dark
soliton is created by employing imaginary time propagation subject
to an enforced axially-symmetric $\pi$ phase step, and then its
dynamics is calculated by using real time propagation.

\section{The soliton periods across the BCS-BEC crossover}
%===========================fig1===============================%
\begin{figure}
\includegraphics[scale=0.3]{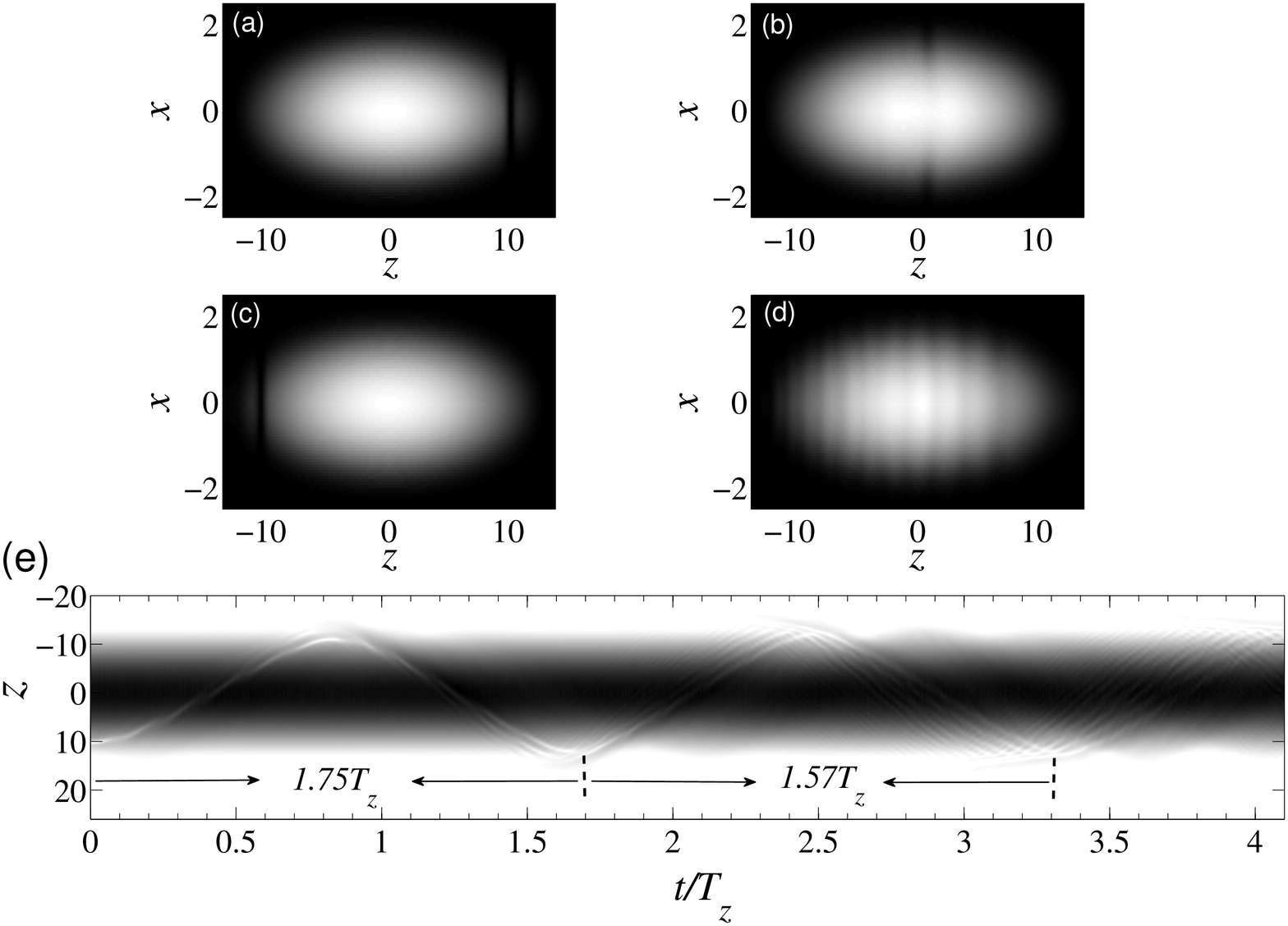}
\caption{\footnotesize{(Color online) Dark solitons dynamics in
anisotropic superfluid Fermi gases in the unitarity limit, with the total
atom pairs $N=2\times10^2$ and aspect ratio $\lambda=6.5$. Plane(a) is
the initial density profiles of dark soliton generated at the axial
position $z_0=3R_z/4$ with $R_z$ being the axial half length. Plane
(b), (c) and (d) are snapshots of dark soliton dynamics at times
$t=0.4T_z$, $t=0.8T_z$, and $t=3.6T_z$, respectively. Light/dark
regions indicate high/low densities. Plane (e) is the spatiotemporal
contour plot of the axial density at $r=0$ plane. The dark soliton
oscillates in the trap with the first period of $1.75T_z$ and the
reduced second of $1.57T_z$. }}
\end{figure}
%===========================fig1===============================%
%%%%%%%
The beginning of our calculation is to apply the parameters
$N=2\times10^2$ and $\lambda=6.5$ for the superfluid Fermi gas in
the unitarity limit($\eta=0$). By imposing $\pi$ phase step at the
axial position $z_0=3R_z/4$ with $R_z$ being the axial half length,
the density profile evolves into that of an axially-symmetric dark
soliton shown in Fig.1(a). This off-center dark soliton under
a harmonic confinement is expected to oscillate back and forth along the
trap as a quasiparticle\cite{bus,vab,hua1}. Fig.1(b)-(d) depict the
density profiles of the dark soliton at different evolution times. As shown in
Fig.1(b), the initial black soliton moves towards the trap center
at $t=0.4T_z$ from the right side, in which it becomes a shallower gray
soliton due to the higher density near the center and the faster it
gets. For the timescale of axial trapping period
$t=0.8T_z$ in Fig.1(c), the soliton is prone to be
instable, emitting radiation in the form of sound waves. After a
long live oscillations accompanied by the sound waves at
$t=3.6T_z$, the dark soliton completely decays into a train
of sound waves(see Fig.1(d)).

In order to monitor the soliton trajectory, we give the
spatiotemporal evolution of density along a cross section at $r=0$
in Fig.1(e), where the soliton and sound waves are indicated by
the light regions. Fig.1(e) provides a clearer indication of
dynamical instability in the form of sound radiation, and gives the first
period of the dark soliton oscillating in the trap to be $T_s=1.75T_z$.
It is shown that the dissipation of energy by the
sound waves from the soliton is associated with increase of the
amplitude of oscillation(anti-damping), and soliton becomes shallower and, as a
result, it accelerates with the second oscillation period of $1.57T_z$.
%%%%%%
%===========================fig2===============================%
\begin{figure}
\includegraphics[scale=0.3]{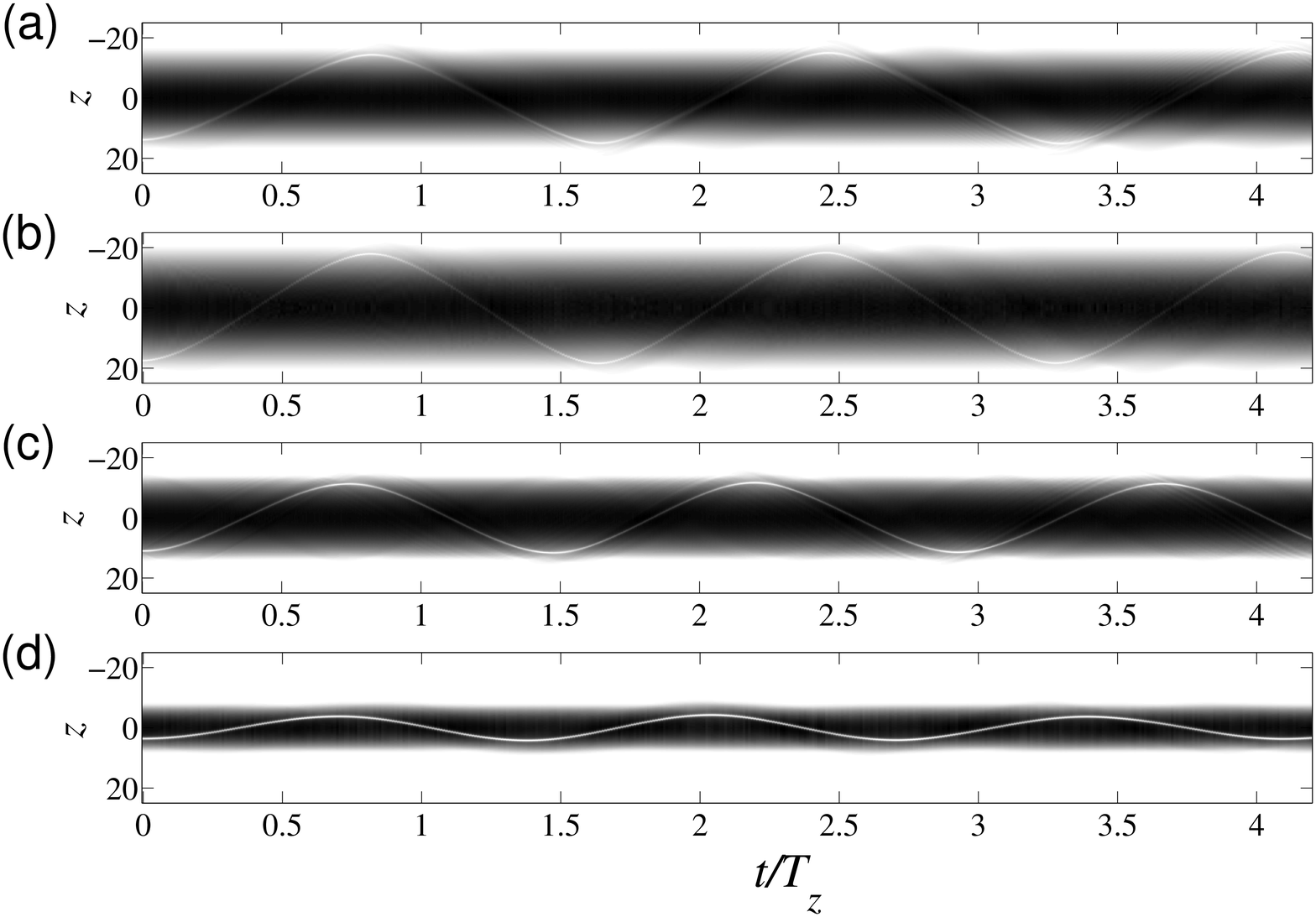}
\caption{\footnotesize{(Color online) The spatiotemporal contour
plots of dark soliton dynamics in superfluid Fermi gases containing
$N=2\times10^2$ atomic pairs. (a) corresponds to $\eta=0$(the
unitarity limit) with the aspect ratio $\lambda=15$, and (b) for
$\eta=0$ with $\lambda=30$; (c) and (d) are for $\eta=1.0$(the BEC
side) and $\eta=6.0$(the BEC limit), respectively, both with
$\lambda=30$.}}
\end{figure}
%===========================fig2===============================%
%%%%%%%

Such dynamical decay of a moving soliton via the emission of sound
waves can be accounted for two instability mechanisms in the
framework of the order-parameter equation: (i) axial background
inhomogeneity due to the trapping potential\cite{hua1,par,bra,dmi},
and (ii) the effects of transverse degree of freedom coupled with
the axial degree by the atomic interaction\cite{hua1,mury,the,wel}.

%%%%%%%%%
%===========================fig3===============================%
\begin{figure}
\includegraphics[scale=0.25]{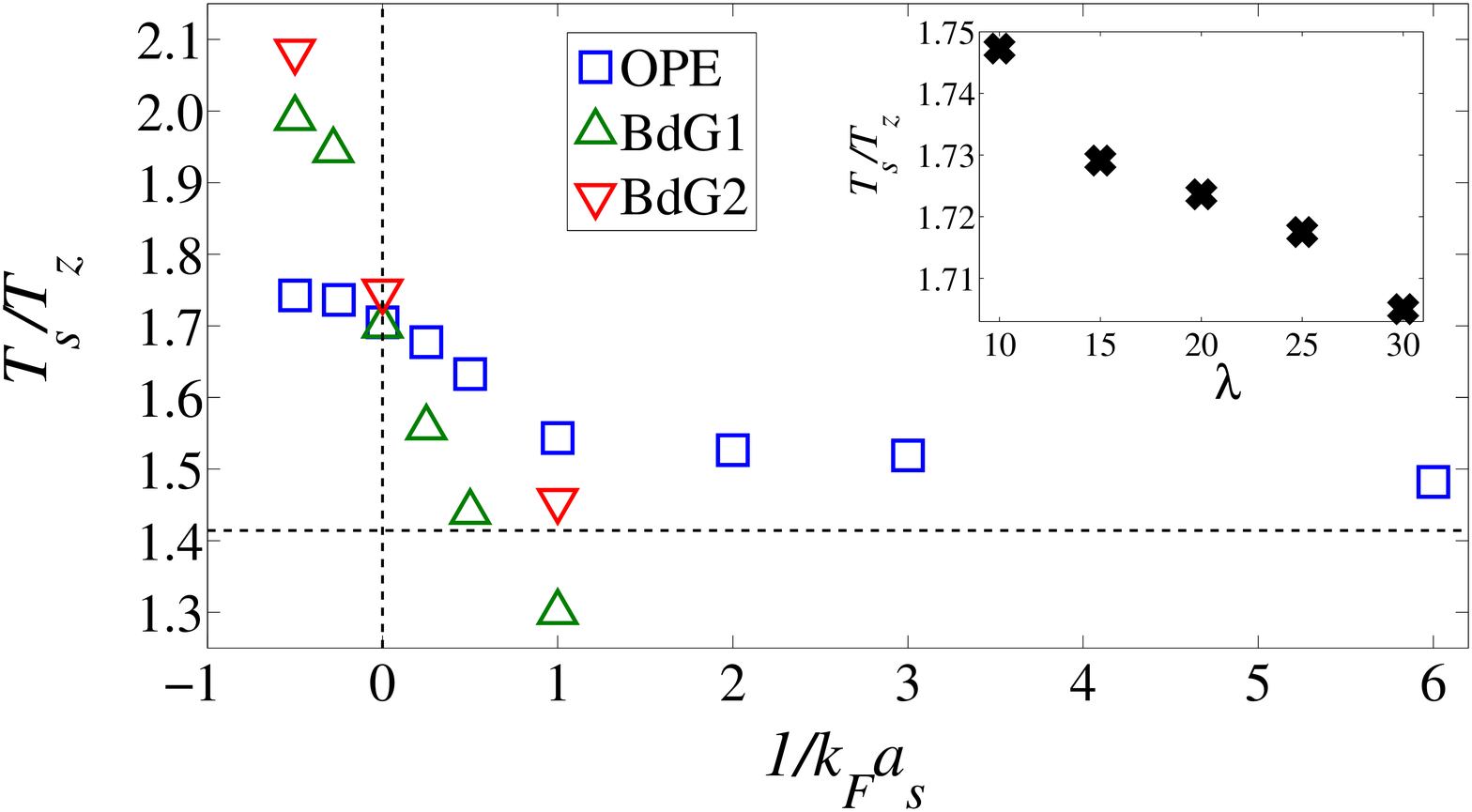}
\caption{\footnotesize{(Color online) Periods of dark solitons in the quasi-1D
regime along the BCS-BEC crossover. Our results($\Box$) based on the
order-parameter equation are compared with ones from the the BdG
equations by Scott {\it et al.}($\bigtriangleup$)\cite{sco} and Liao
{\it et al.}($\bigtriangledown$)\cite{ren}. The horizontal dashed
line indicates the well-known period $\sqrt{2}T_z$ for atomic BECs.
The inset shows the periods of the soliton oscillating in the unitarity
limit as a function of the aspect ratio. }}
\end{figure}
%===========================fig3===============================%
%%%%%%%

The relevant size of solitons in the crossover from the BEC limit up
to the unitary limit can be characterized by the healing length
$\xi=\hbar/\sqrt{2M\mu_s}$, with local chemical potential
$\mu_s({\bf r})$ depending on the spatially inhomogeneous density
$n_s({\bf r})$. The local chemical potential is determined by the
ground state solution of the order-parameter equation Eq.(\ref{gg}),
that is
$\mu_s({\bf r})+V_s({\bf r})=\mu_G$.
By using the normalized condition, one can obtain the bulk chemical
potential $\mu_G$\cite{wen1}. In the presence of the harmonic trap,
the axial size of the superfluid $R_z=\sqrt{2\mu_G/M\omega^2_z}$ is set
by the trapping frequency. For the case of Fig.1, we find that
$R_z/\xi=2\sqrt{\mu_s\mu_G}/\hbar\omega_z=63$, which
means that the change in the axial background is very weak over the
size of soliton.

On the other hand, the characteristic size of the soliton is the
order of the healing length, thus the corresponding (axial) kinetic
energy is the order of $\mu_s$. The dimensionality parameter defined by
$\alpha=\mu_s/\hbar\omega_{\perp}=3.22$ implies that the kinetic
energy is larger than the transverse energy $\hbar\omega_{\perp}$,
and the atomic interaction can induce transfer of axial energy to
the transversal degrees of freedom. Therefore, we conclude that the
dominant decay mechanism is due to the coupling to transverse modes.

To suppress the dynamical instability, we perform a stronger
transverse confinement of their motion, that is $\lambda=15$(
$\alpha=2.44$ correspondingly).  Therefore, the growth of transverse
energy $\hbar\omega_{\perp}$ suppresses the transfer of kinetic energy of
the soliton to transverse modes and resulting in the decay of the
soliton(see Fig.2(a)). For a tight enough trapping
potential($\lambda=30,\alpha=1.93$) in Fig.2(b), the instability does not occur
as a consequence of possibility of separation of the axial and transversal
degrees of the freedom.  It is seen that the period of the stable soliton oscillation is
$T_s=1.7T_z$, which agrees well with one
calculated by the BdG equations\cite{sco,ren} for the case in the unitarity limit.
As shown in the inset of Fig.3, the soliton periods have a relatively weak dependence on
the anisotropy, that is larger values of $\lambda$ yield smaller
periods.  Notice that under the conditions of tight transverse
confinement $\lambda=30$ and small pairs number $N=2\times10^2$, the
system is in the quasi-1D regime that corresponds to the cases discussed
by the mean-field theory\cite{sco,ren}. We also present the
spatiotemporal evolutions of dark solitons in the quasi-1D regime for the cases of the
BEC side in Fig.2(c) and BEC limit in Fig.2(d), respectively.

Fig.3 shows our results ($\Box$) on the periods $T_s$ of dark
soliton in the quasi-1D regime as a function of the dimensionless
interaction parameter $\eta\equiv1/k_Fa_s$. The decrease of $T_s$ as
one moves from the BCS regime to the BEC limit is consistent with
the mean-field theory computations by Scott {\it et
al.}($\bigtriangleup$)\cite{sco} and Liao {\it et
al.}($\bigtriangledown$)\cite{ren}. In the BEC limit($\eta=6$), our
result is very close to the well-known value $\sqrt{2}T_z$ for
atomic BECs\cite{hua1,bus,vab}, which is indicated by the horizontal
dashed line in Fig.3. We find that in the BEC regime($\eta>0$), our
results are slightly larger than ones based on the extended BdG equations. It is
because that the BdG equations can not obtain the beyond mean-field term of the equation of state
correctly¡£\cite{gio}. Notice that
the order-parameter equation fails to give correct results in the
BCS side. For comparison purposes, we also present our results in the BCS side($\eta<0$).
Interestingly, different from the small discrepancy in the BEC
regime, in the BCS side the BdG results are significantly larger than our results,
which may be interpreted by the coupling to fermionic quasiparticle excitations
near the soliton\cite{sco} that are completely disregarded in the order-parameter equation.

\section{Snake instability of unitary Fermi gases}

Dark solitons have 1D character, which are stable in the quasi-1D regime, but
feature a long-wavelength transverse instability known as the "snake
instability"\cite{hua2,mur,fed,jbr,pgk}, when extended into higher
dimensions. The snake instability originates from the
transfer of the soliton kinetic energy to the transverse modes
parallel to the soliton nodal plane. Generally dark solitons undergo a
snake deformation, causing the nodal plane collapse into vortex rings\cite{fed,jbr,pgk} in
3D (or vortex-antivortex pairs in two
dimensions\cite{hua2}).  For atomic BECs, it has been shown
that this instability leads to a strong bending of the nodal plane,
which breaks down into vortex rings and sound waves, as
experimentally observed\cite{dut,bpa}.
%%%%%%
%===========================fig4===============================%
\begin{figure}
\includegraphics[scale=0.3]{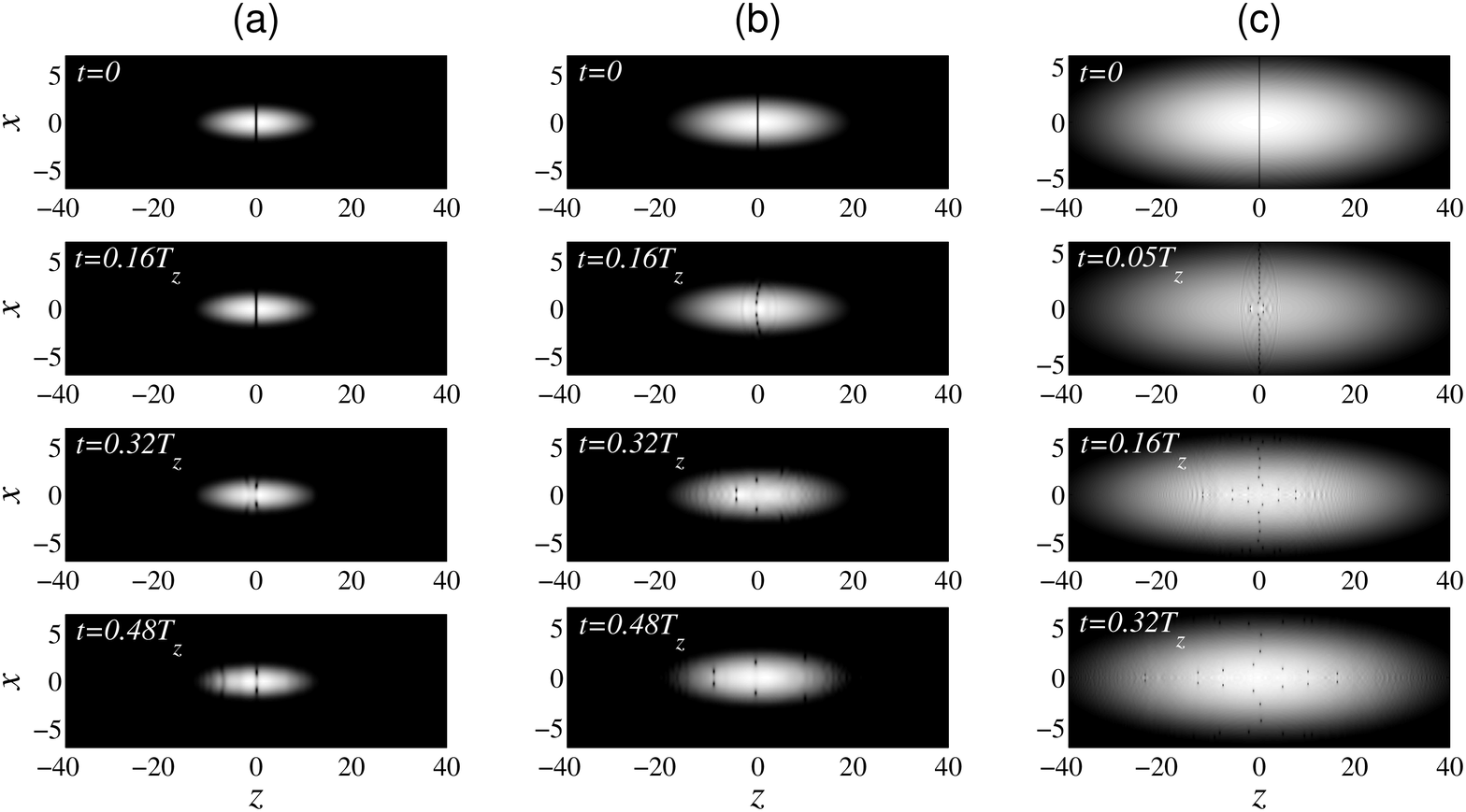}
\caption{\footnotesize{(Color online) The snapshots of the
stationary dark soliton dynamics in the unitarity limit with $\lambda=6.5$.
Planes (a), (b) and (c) correspond to the total numbers of atom pairs
$N=2\times10^2$, $N=2\times10^3$ and $N=2\times10^5$, respectively.}}
\end{figure}
%===========================fig4he===============================%
%%%%%%%

The snake stability of dark solitons of superfluid Fermi gases in
the unitarity limit can be studied by monitoring the evolution of a standing
dark soliton created at the trap center\cite{hua2,mur,fed,jbr,pgk}. The development of the snake
instability and the concomitant vortex rings for a wide range of the numbers of atom pairs with
$\lambda=6.5$ are displayed in Fig.4(a) for $N=2\times10^2$,
Fig.4(b) for $2\times10^3$, and Fig.4(c) for $2\times10^5$,
respectively. Vortex rings resemble toroids where the superfluid
density is depleted, and so the slice of the vortex ring appears as
two dark spots separated vertically. As shown in
Fig.4(a), the stationary dark soliton is subject to the snake
instability, bending into one vortex ring, in strongly contrast to the moving
soliton slowly decaying into sound radiation(see Fig.1). This is due to the
larger dimensionality parameter $\alpha=\mu_G/\hbar\omega_{\perp}=7.38$ at the
trap center. As the number of atom pairs rises, the increased chemical potential
opens more decay channels, which result in the formation
of more vortex rings, that is 3 [$\alpha$=15.90] in Fig.4(b) and
14[$\alpha$=73.80] in Fig.4(c). For comparison purposes, the results
for the three cases are presented on the same spatial grid. From
plane (a) to (c), the chemical potential increases tenfold and
inversely the size of solitons and vortex rings decreases
threefold. We can estimate the size of the nonlinear excitations
for the experimental relevant case of $N=2\times10^5$ in Fig.4(c). The
size scale is given by $\xi=R_z/959=0.29$$\rm \mu m$, which is too small
to be resolved by optical mean directly, but by the time of flight expansion
acting as a magnifying glass\cite{yef}. In addition, we
find the time for the start of snake instability reduces
significantly from $0.32T_z$ in Fig.4(a) to $0.05T_z$ in Fig.4(c).

Now we show that how the instability mechanism can be suppressed under
tight transverse confinements\cite{npp} and determine the criterion for stability against the
transverse decay\cite{mury}. We consider the superfluid Fermi gases containing $N=2\times10^4$ atom
pairs in the unitarity limit, and examine the evolution of dark solitons generated at an
off-center position of $3R_z/4$. Fig.5 shows the density(left planes) and corresponding
phase(right planes) of the evolutions of the solitons at the time when the solitons reach
the trap center. Plane (a), (b), (c), (d), and (e) correspond to $\lambda=6.5$, 50, 100, 180 and 250,
and the evolution time $t=0.56T_z$, $0.53T_z$, $0.47T_z$, $0.45T_z$, and $0.42T_z$, respectively.

With a weak transverse trapping $\lambda=6.5$ in Fig.5(a),
the created soliton subjecting snake instability is decay to
vortex rings, one of which reaches the trap center at $t=0.56T_z$. Vortex rings
can be evidenced by the $2\pi$ phase change at
any point of the circle, as shown in the right plane of Fig.5(a). Increasing the
transverse frequency $\lambda=50$ in Fig.5(b) leads to a decrease in the bending of
the soliton, and hence the production of a single vortex ring. It is seen that only by observing dips in the density
profile from the left plane of Fig.5(b), it is very hard to discriminate between soliton and vortex rings.
We find that the vortex ring evolves back into a soliton, when moving near the ends of the trap due to the decrease of the dimensionality parameter.
It is seen that such periodic soliton/vortex ring is stable with a oscillation period
of $2.2T_z$, which was observed firstly in atomic BECs\cite{sk}. In the geometries($\lambda=100$ in Fig.5(c) and $\lambda=180$
in Fig.5(d)) where the soliton is transversely unstable but the transverse width of the system is too small to support vortex rings,
an excitation with soliton and vortex properties, known as a hybrid of soliton and vortex rings\cite{sk,sho},
is predicted to occur. The hybrid of solitons and vortex rings can be evidenced in the right planes,
by the emergence of not only phase azimuthal dependence, but also phase jump that is
the character of soliton. We find that the oscillation period of the hybrid of soliton
and vortex rings is $2.0T_z$ for the case of Fig.5(c).  Finally, very tight transverse confinement($\lambda=250$)
results in a highly elongated quasi-1D geometry, evidencing by a stable soliton with its step phase profile
of Fig.5(e).  The  period of the stable soliton in such highly elongated geometry
is found to be $1.83T_z$, only 8\% larger than one in the quasi-1D regime.
%%%%%%
%===========================fig5===============================%
\begin{figure}
\includegraphics[scale=0.3]{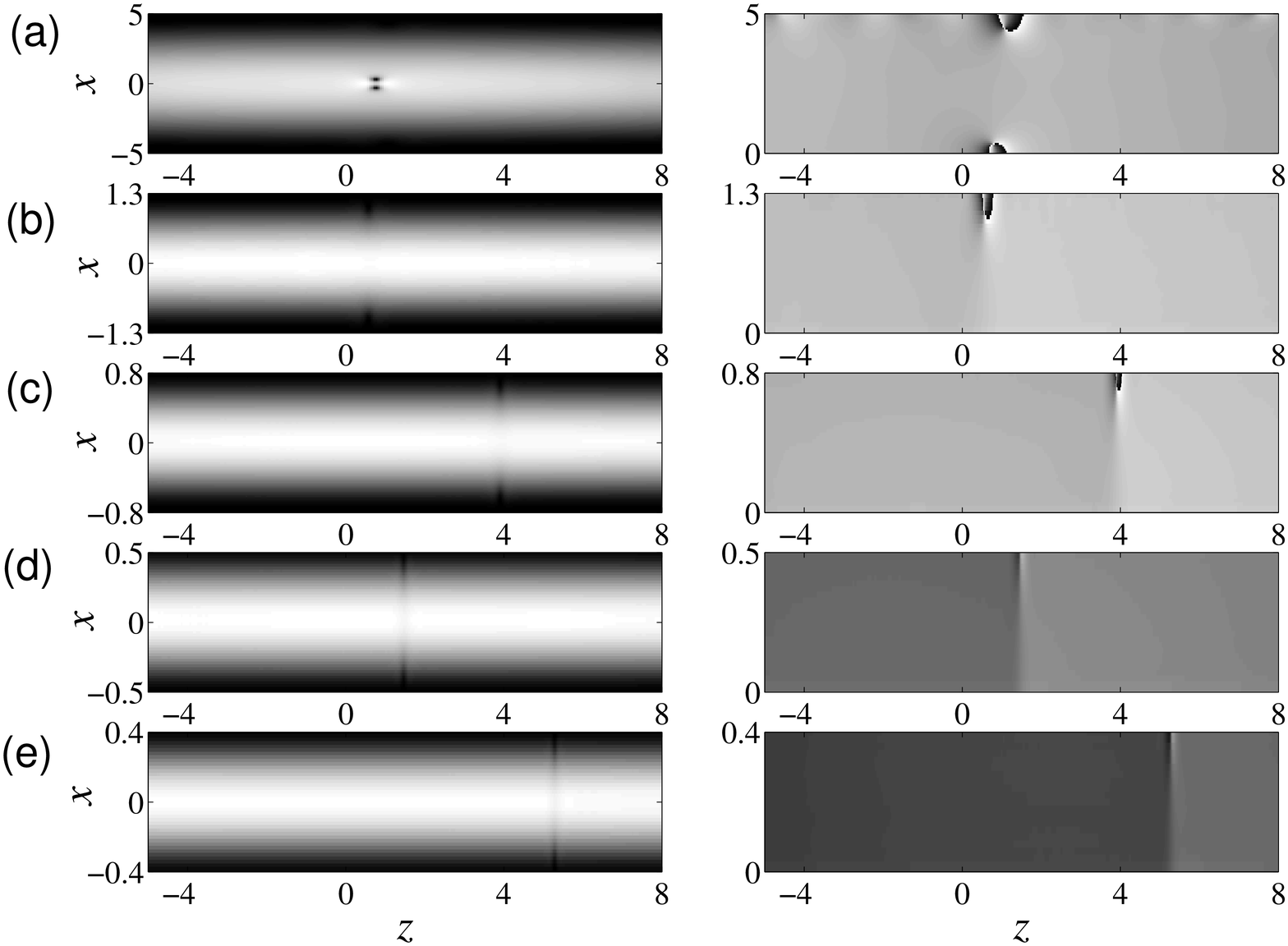}
\caption{\footnotesize{(Color online) Close-up snapshots of density(left) and phase(right) profiles
of the evolutions of the off-centered dark solitons initially generated at $3R_z/4$, when they evolves
near the center of the unitary Fermi gases with $N=2\times10^4$ for different transverse confinement.
Plane(a), (b), (c), (d), and (e) correspond to $\lambda=6.5$, 50, 100, 180 and 250,
and  evolution time $t=0.56T_z$, $0.53T_z$, $0.47T_z$, $0.45T_z$, and $0.42T_z$, respectively. }}
\end{figure}
%===========================fig5===============================%
%%%%%%%

Therefore, the criterion of dynamical stability of dark solitons in trapped unitary Fermi gases
can be estimated\cite{mury} by the case of Fig.5(e), that is $\alpha_c=\mu_s/\hbar\omega_{\perp}=4.2$.
We find that the stability criterion, that is $\alpha<\alpha_c$, is very strict for the conditions of current
experiments. For the system with total number of atom pairs of $10^5$ in most Fermi experiment, it
requires at least $\lambda=1500$ in the unitarity limit when solitons are generated in the off-center position of
$3R_z/4$, while only $\lambda=350$ for $\alpha_c=2.4$ of weakly interacting atomic BECs\cite{mur}.

\section{comparison with experiment}
%%%%%%
%===========================fig6===============================%
\begin{figure}
\includegraphics[scale=0.3]{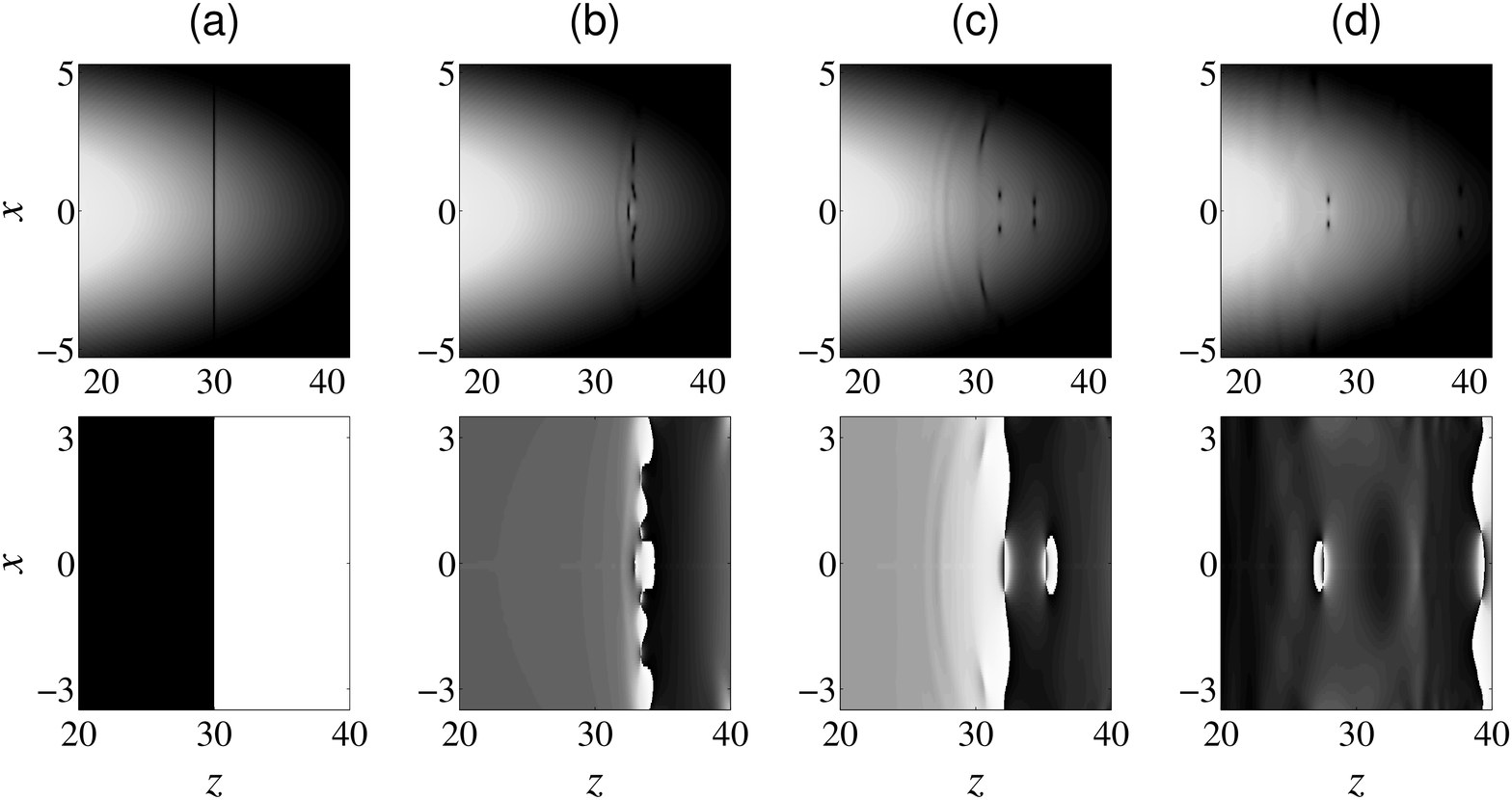}
\caption{\footnotesize{(Color online) Close-up snapshots of the evolution of
density(top) and corresponding phase profiles(bottom) for the dark
dynamics in the unitary Fermi gas with the MIT experimental
parameters of $N=2\times10^5$ and $\lambda=6.5$ at (a)$t=0$,
(b)$t=0.07T_z$, (c)$t=0.12T_z$ and (d)$t=0.24T_z$, showing the onset
of the snake instability and the
decay of the soliton into two vortex rings as evidenced by the
corresponding phase profiles.}}
\end{figure}
%===========================fig6===============================%
%%%%%%%
In the MIT experiment\cite{yef}, the superfluid Fermi gas containing
$2\times10^5$ atom pairs was prepared in a cylindrically symmetric trap with
$\omega_z=2\pi\times10.66$$\rm Hz$ and $\lambda=6.5$. In order to
observe dark solitons dynamics in the Fermi gas, they optically applied a step-function
potential to advance a $\pi$ phase shift of the superfluid order parameter, thereby
imprinting a moving soliton at the off-center position.

We perform numerical simulations using the experimental parameters in the unitary limit.
The results of the close-up snapshots of density profiles(top row) of dark soliton dynamics and
corresponding phase profiles(bottom row) are presented in Fig.6.
In Fig.6(a), the initial soliton has a node of zero density at $3R_z/4$(top) and
a $\pi$ phase step(bottom). As the soliton starts to move at $t=0.07T_z$(6.56{\rm ms}), the
soliton plane is dynamically unstable subjecting to a gradual bending shown in Fig.6(b),
which is resulted from the inhomogeneous transverse density. Subsequently in Fig.6(c), the soliton plane
tears into pieces, creating two vortex rings and radiating sound waves. Vortex ring is evidenced by a $2\pi$ phase
dependence azimuthally in the bottom plane. At $11.2{\rm ms}$(Fig.6(c)), the produced two vortex rings propagate
in opposing directions, that is one propagates to the left and another to the right. Finally the vortex
ring propagating rightly is absorbed by the boundary, and only the left vortex ring survives.
Note that the dimensionality parameter at $3R_z/4$ position is $\alpha=32$, much larger than
the stability criterion $\alpha_c=4.2$ estimated by us, so such decay channel can be anticipated.

%%%%%%
%===========================fig7===============================%
\begin{figure}
\includegraphics[scale=0.3]{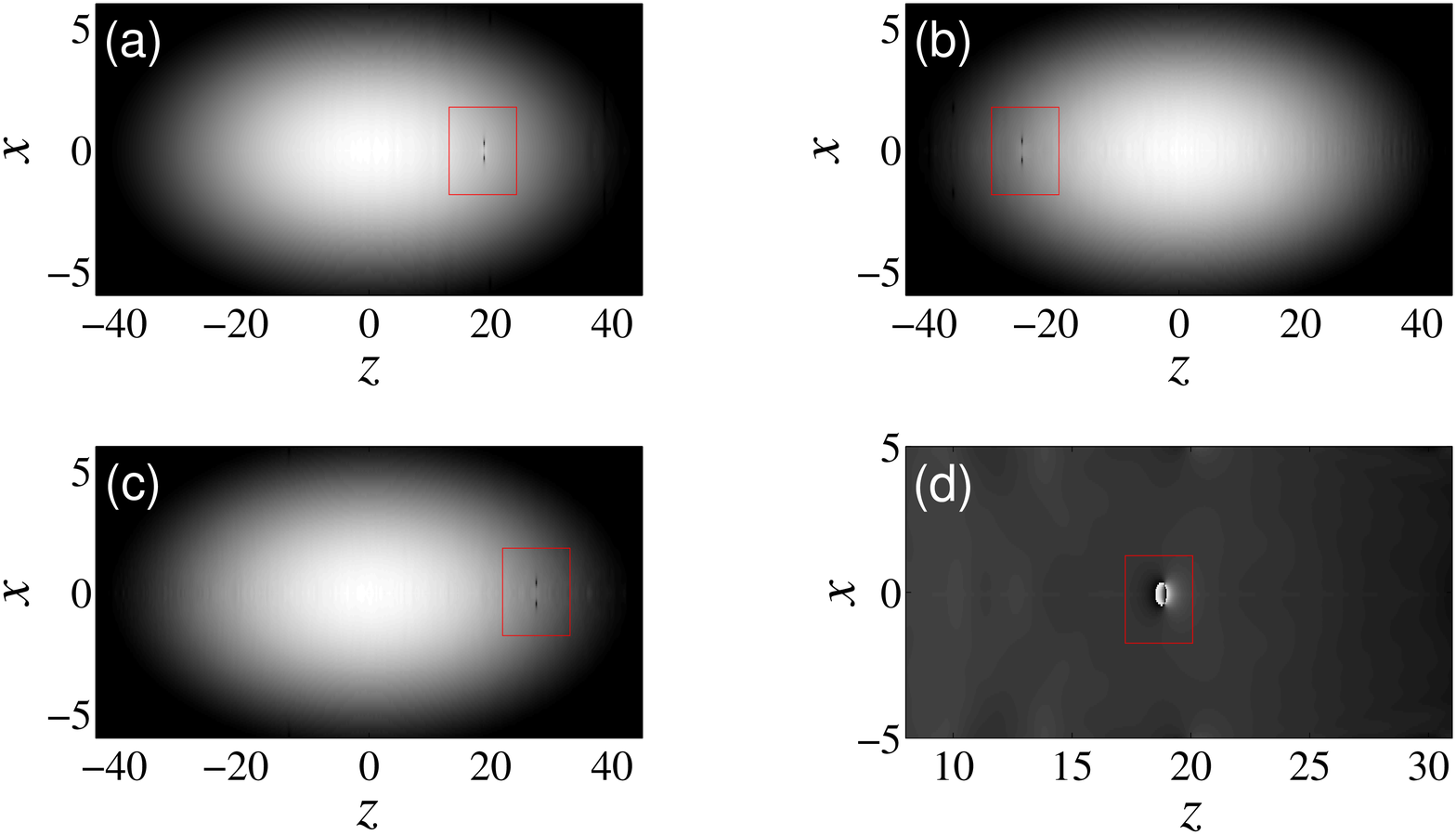}
\caption{\footnotesize{(Color online) Snapshots of the oscillation of a vortex ring
created by the snake instability of the unitary Fermi gases from
Fig.6 at (a) $t=0.4T_z$ as well as (d) corresponding phase,
(b)$t=2.0T_z$, and (c)$t=2.8T_z$.}}
\end{figure}
%===========================fig7===============================%
%%%%%%%

It is interesting to investigate the evolution of the survival vortex ring in the trapped Fermi gas.
As shown in Fig.7, we find that the vortex ring is stable and presents a oscillation in
a soliton-like manner, which is highlighted by red lines. Interestingly, after reflecting by the left end, it is also
disappear when moves to the right boundary, thus only a single oscillation period can be observed.
This period is given by about $2.8T_z$ that is nearly two times than the soliton period of $1.7T_z$,
and also different from the period of $2.0T_z$ for a hybrid of solitons and vortex rings.
The results suggest that in a strongly interacting superfluidity
to probe 1D character of
stable solitons needs very small number of atoms or very large harmonic trapping frequency against the snake instability,
which is hard to be  fulfilled by current experimental situation. But it may be an ideal system to
experimentally study the dynamics of vortex rings or hybrids of solitons and vortex rings resulting from
the snake instability, which
can be distinguished by measuring different oscillation periods or their phases profiles as we discussed.

%%%%%%%%%%%%%%%%%%%%%%%%%%%%%%%%%%%%%%%%%%%%%%%%%%%%%%%%%%%%%%%%%
%%%%%%%%%%%%%%%%%%%%%%%%%%%%%%%%%%%%%%%%%%%%%%%%%%%%%%%%%%%%%%%%%

%%%%%%%%%%%%%%%%%%%%%%%%%%%%%%%%%%%%%%%%%%%%%%%%%%%%%%%%%%%%%%%%%
\section{Conclusions}
We perform the calculations for the dynamics and stability of dark solitons in anisotropic
superfluid Fermi gases for a wide range of atomic particle numbers and ratio
aspects within the framework of the order-parameter equation. We study the dynamics of
solitons in the trapped superfluid Fermi gases with small number of atoms, and the
computed soliton  period of $1.7T_z$ in the unitary limit is in a good agreement with one by
the BdG equations.  The snake instability of unitary Fermi gases is studied.
By examining the evolutions of an initially off-center dark soliton under various aspect ratios,
and a transition from hybrids of solitons and vortex rings to stable soltion discriminated by their phase profiles,
we give the criterion for stability against transverse decay, that is $\alpha_c=4.2$. In
addition, it is found that the soliton period increases 8\% as the number of atom pairs is increased two orders,
and the hybrid of solitons and vortex rings has a larger period of $2.0T_z$.
We simulate the recent MIT experiment on the dark soliton dynamics in the unitary Fermi gas. Instead of performing
a very slow oscillation as observed experimentally, the imprinted soliton is found to evolve into vortex rings,
which propagate in soliton-like manner with a period of $2.8T_z$. Such disagreement between theory and experiment may be accounted for the time-of-flight method with the Feshbach resonance, which will be considered in the future.

After finishing this manuscript,
we became aware of Ref.\cite{bul1} and Ref.\cite{cet} addressing the
snake instability of the unitary Fermi gases using different theoretical frameworks, which reached the same conclusions.

%%%%%%%%%%%%%

\acknowledgments

We gratefully acknowledge Renyuan Liao and Hui Zhai for enlightening discussions and helpful comments.
W.W. is supported by the NSFC under Grant No. 11105039, the Fundamental Research Funds for Central Universities
of China (program No. 2012B05714 and 2010B23414), and Doctoral Foundation of Hohai university 2010. X.D.M. is
supported by the NSFC Grant No. 11264039 and the Key Research Project of Xinjiang Higher Education, China under
Grant No. XJED2010141.

%%%%%%%%%%%%%%%%%%%%%%%%%%%%%%%%%%%%%%%%%%%%%%%%%%%%%%%%%%%%%%

%%%%%%%%%%%%%%%%%%%%%%%%%%%%%%%%%%%%%%%%%%%%%%%%%%%%%%%%%%%%%%


\begin{references}
%soliton book
\bibitem{dau} T. Dauxois and M. Peyrard, {\it Physics of Solitons} (Cambridge University Press,
Cambridge, 2006).

\bibitem{pet} C. J. Pethick and H. Smith, {\it Bose-Einstein condensation in Dilute
Gases}, 2nd edn. (Cambridge University Press, Cambridge, 2008).

\bibitem{kev} P. G. Kevrekidis, D. J. Frantzeskakis, and R.
Carretero-Gonz{\'a}lez, ed. {\it Emergent Nonlinear Phenomena in
Bose-Einstein Condensates} (Springer-Verlag, Berlin, 2008).

\bibitem{fra} D. J. Frantzeskakis, J. Phys. A:Math. Theor. {\bf 43},
213001 (2010).

%SFs firstly experimental realizations
\bibitem{hara} K. M. O'Hara, S. L. Hemmer, M. E. Gehm, S. R. Granade, and J. E. Thomas,
Science {\bf 298}, 2197 (2002).

\bibitem{reg}C. A. Regal, M. Greiner and D. S. Jin, Phys. Rev. Lett. {\bf 92},
040403 (2004).

\bibitem{abr} M. Bartenstein, A. Altmeyer, S. Riedl, S. Jochim, C. Chin, J. H. Denschlag,
and R. Grimm, Phys. Rev. Lett. {\bf 92}, 120401 (2004).


\bibitem{zwi} M. W. Zwierlein, C. A. Stan, C. H. Schunck, S. M. F. Raupach, A. J. Kerman,
and W. Ketterle, Phys. Rev. Lett. {\bf 92}, 120403 (2004).


%SFs reviews
\bibitem{gio}S. Giorgini, L. P. Pitaevskii, and S. Stringari, Rev. Mod. Phys. {\bf 80}, 1215 (2008), and references therein.

%new review on quantum gases
\bibitem{ada}A. Adams, L. D. Carr, T. Sch$\ddot{a}$fer, P. Steinberg
and J. E. Thomas, New J. Phys. {\bf 14}, 115009 (2012).

%soliton in SFS
\bibitem{yef} T. Yefsah, A. T. Sommer, M. J. H. Ku, L. W. Cheuk,
W. Ji, W. S. Bakr, and M. W. Zwierlein, Nature {\bf 499}, 426
(2013).

%Extended BCS method for soliton
\bibitem{ant} M. Antezza, F. Dalfovo, L. P. Pitaevskii, and S.
Stringari, Phys. Rev. A {\bf 76}, 043610 (2007).

\bibitem{spu} A. Spuntarelli, L. D. Carr, P. Pieri, and G. C.
Strinati, New J. Phys. {\bf 13}, 035010 (2011).

\bibitem{sco} R. G. Scott, F. Dalfovo, L. P. Pitaevskii, and S.
Stringari, Phys. Rev. Lett. {\bf 106}, 185301 (2011).

\bibitem{ren} R. Liao and J. Brand, Phys. Rev. A {\bf 83}, 041604(R)
(2011).

\bibitem{sco1} R. G. Scott, F. Dalfovo, L. P. Pitaevskii, S.
Stringari, O. Fialko, R. Liao and J. Brand, New. J. Phys. {\bf 14},
023044 (2012).

%density functional for soliton
\bibitem{bul} A. Bulgac, Y. -L. Luo, and K. J. Roche, Phys. Rev. Lett.
{\bf 108}, 150401 (2012).

%order-parameter method of soliton
\bibitem{wen2} W. Wen and  G. X. Huang, Phys. Rev. A {\bf 79}, 023605 (2009).

\bibitem{khan} A. Khan and P. K. Panigrahi, J. Phys. B: At. Mol. Opt. Phys.
 {\bf 46}, 115302 (2013).


%book for BCS mean-field theory
\bibitem{pdg} P. G. de Gennes, {\it Superconductivity of Metals and
Alloys} (Addison-Wesley, New York, 1989).

\bibitem{sal} L. Salasnich, N. Manini,  and F. Toigo, Phys. Rev. A {\bf 77},
043609 (2008); L. Salasnich and F. Toigo, {\it ibid.} {\bf 78},
053626 (2008).

\bibitem{wen3}W. Wen, Y. Zhou, and G. X. Huang, Phys. Rev. A {\bf 77}, 033623
(2008).

%order parameter equation
\bibitem{kim}Y. E. Kim and A. L. Zubarev, Phys. Rev. A {\bf 70},
033612 (2004).

\bibitem{zub} A. L. Zubarev, J. Phys. B: At. Mol. Opt. Phys. {\bf
42}, 011001 (2009).

\bibitem{adh1} S. K. Adhikari, Phys. Rev. A {\bf 77}, 045602 (2008).

\bibitem{anc} F. Ancilotto, L. Salasnich and F. Toigo, Phys. Rev. A {\bf 85},
063612 (2012).

\bibitem{adh} S. K. Adhikari, J. Phys. B: At. Mol. Opt. Phys.
 {\bf 43}, 085304 (2010).

\bibitem{rup} G. Rupak and T. Sch$\ddot{a}$fer, Nucl. Phys. A {\bf
816}, 52 (2009).

\bibitem{xio} H. W. Xiong, S. J. Liu and M. S. Zhan, Phys.
Rev. A {\bf 74}, 033602 (2006).

%mental carlo
\bibitem{geas} G. E. Astrakharchik, J. Boronat, J. Casulleras, and S. Giorgini,
Phys. Rev. Lett. {\bf 93}, 200404 (2004).


%gamma index
\bibitem{mann} N. Manini and L. Salasnich, Phys. Rev. A {\bf 71}, 033625 (2005).

\bibitem{dia} G. Diana, N. Manini, and L. Salasnich, Phys. Rev. A {\bf
73}, 065601 (2006).

\bibitem{wen1} W. Wen, S.-Q. Shen,  and G. X. Huang, Phys. Rev. B {\bf 81}, 014528 (2010).

\bibitem{noz} P. Nozi\`{e}res and S. Schmitt-Rink, J. Low Temp.
Phys. {\bf 59}, 195 (1985).

\bibitem{pie} P. Pieri and G. C. Strinati, Phys. Rev. Lett. {\bf
91}, 030401 (2003).

%the valid of OPEs
\bibitem{lsa} L. Salasnich, F. Ancilotto, N. Manini, and F. Toigo,
Laser Phys. {\bf 19}, 636 (2009).

\bibitem{anci} F. Ancilotto, L. Salasnich and F. Toigo, Phys. Rev. A
{\bf 79}, 033627 (2009).

%pairs breaking
\bibitem{zai} H. Zhai and T.-L. Ho, Phys. Rev. Lett. {\bf 97}, 180414 (2006).


%phase imprinting
\bibitem{dob} {\L}. Dobrek, M. Gajda, M. Lewenstein, K. Sengstock, G. Birkl and
W. Ertmer, Phys. Rev. A {\bf 60}, R3381 (1999).

\bibitem{bur} S. Burger, K. Bongs, S. Dettmer, W. Ertmer, K. Sengstock, A. Sanpera,
G. V. Shlyapnikov, and M. Lewenstein, Phys. Rev. Lett. {\bf 83},
5198 (1999).

\bibitem{den} J. Denschlag, J. E. Simsarian, D. L. Feder, C. W. Clark, L. A. Collins,
J. Cubizolles, L. Deng, E. W. Hagley, K. Helmerson, W. P. Reinhardt,
S. L. Rolston, B. I. Schneider, and W. D. Phillips,  Science {\bf 287},
97 (2000).


%numerical method
\bibitem{pmu} P. Muruganandam and S. K. Adhikari, Comput. Phys.
Commun. {\bf 180}, 1888 (2009).

%oscillation of soliton
\bibitem{bus} Th. Busch and J. R. Anglin, Phys. Rev. Lett. {\bf 84},
2298 (2000).

\bibitem{hua1} G. X. Huang, J. Szeftel and S. H. Zhu, Phys. Rev. A
{\bf 65}, 053605 (2002).

\bibitem{vab} V. A. Brazhnyi, V. V. Konotop, and L. P. Pitaevskii,
Phys. Rev. A {\bf 73}, 053601 (2006).

%sound emitting


%axial imhomogenity
\bibitem{par} N. G. Parker, N. P. Proukakis, M. Leadbeater, and C.
S. Adams, Phys. Rev. Lett. {\bf 90}, 220401 (2003); N. G. Parker, N.
P. Proukakis, M. Leadbeater and C. S. Adams, J. Phys. B: At. Mol.
Opt. Phys. {\bf 36}, 2891 (2003).


\bibitem{bra} V. V. Konotop and L. Pitaevskii, Phys. Rev. Lett. {\bf
93}, 240403 (2004);  V. A. Brazhnyi and V. V. Konotop, Phys. Rev. A
{\bf 68}, 043613 (2003).

\bibitem{dmi} D. E. Pelinovsky, D. J. Frantzeskakis, and P. G.
Kevrekidis, Phys. Rev. E {\bf 72}, 016615 (2005).


%coupling with transversal modes

\bibitem{mury} A. Muryshev, G. V. Shlyapnikov, W. Ertmer, K.
Sengstock, and M. Lewenstein, Phys. Rev. Lett. {\bf 89}, 110401
(2002).

\bibitem{the} G. Theocharis, P. G. Kevrekidis, M. K. Oberthaler, and
D. J. Frantzeskakis, Phys. Rev. A {\bf 76}, 045601 (2007).


%higher dimensional instability

\bibitem{wel} A. Weller, J. P. Ronzheimer, C. Gross, J. Esteve,
M. K. Oberthaler, D. J. Frantzeskakis, G. Theocharis and P. G.
Kevrekidis, Phys. Rev. Lett. {\bf 101}, 130401 (2008).


% the snaking instabilities of solitons in BEC

\bibitem{mur} A. E. Muryshev, H. B. van Linden van den Heuvell, and
G. V. Shlyapnikov, Phys. Rev. A {\bf 60}, R2665 (1999).

\bibitem{hua2} G. X. Huang, V. A. Makarov and M. G. Velarde, Phys. Rev. A
{\bf 67}, 023604 (2003).

\bibitem{fed} D. L. Feder, M. S. Pindzola, L. A. Collins, B. I.
Schneider, and C. W. Clark, Phys. Rev. A {\bf 62}, 053606 (2000).

\bibitem{jbr} J. Brand and W. P. Reinhardt, Phys. Rev. A {\bf 65},
043612 (2002).

\bibitem{pgk} P. G. Kevrekidis, G. Theocharis, D. J. Frantzeskakis, and A. Trombettoni,
Phys. Rev. A {\bf 70}, 023602 (2004).

% experimental vortex rings
\bibitem{dut} Z. Dutton, M. Budde, C. Slowe, and L. V. Hau, Science
{\bf 293}, 663 (2001).

\bibitem{bpa} B. P. Anderson, P. C. Haljan, C. A. Regal, D. L. Feder, L.
A. Collins, C. W. Clark, and E. A. Cornell, Phys. Rev. Lett. {\bf
86}, 2926 (2001).

%stability criterion
\bibitem{npp} N. P. Proukakis, N. G. Parker, D. J. Frantzeskakis and
C. S. Adams, J. Opt. B: Quantum Semiclass. Opt. {\bf 6}, S380 (2004).

%hybrid of soliton and vortex rings
\bibitem{sk} S. Komineas and N. Papanicolaou, Phys. Rev. Lett. {\bf
89}, 070402 (2002); S. Komineas and N. Papanicolaou, Phys. Rev. A {\bf 67},
023615 (2003).

\bibitem{sho} I. Shomroni, E. Lahoud, S. Levy and J. Steinhauer, Nature Phys. {\bf 5},
193 (2009).

% recent notation
\bibitem{bul1} A. Bulgac, M. M. Forbes, M. M. Kelley, K. J. Roche, and G. Wlaz{\l}owski,
e-print arXiv:1306.4266.

\bibitem{cet} A. Cetoli, J. Brand, R. G. Scott, F. Dalfovo, and L. P. Pitaevskii,
e-print arXiv:1307.3717.













%%%%%%%%%%%%%%%%%%%%%%%%%%%%%%%%%%%%%%%%%%%%%%%%%%%%%%%%%%%%%%%%%


\end{references}
\end{document}